\documentclass[conference,compsoc]{IEEEtran}
\usepackage{graphicx}
\usepackage{acronym}
\usepackage{balance}
\usepackage{graphicx}
\usepackage{multirow}
\usepackage{multicol}
\usepackage{algorithm}
\usepackage{algorithmicx}
\usepackage{algpseudocode}
\usepackage{xspace}
\usepackage{subfigure}
\usepackage{pifont}
\usepackage{booktabs}
\usepackage{amssymb}
\usepackage{amsmath}
\usepackage{mathtools}
\usepackage{wrapfig}
\usepackage{hyperref}
\usepackage{xcolor}
\usepackage{balance}
\newcommand{\cmark}{\ding{51}}%
\newcommand{\xmark}{\ding{55}}%
\usepackage[paperheight=11in,paperwidth=8.5in,left=1.62cm,right=1.62cm,top=1.9cm]{geometry}

\acrodef{MEC}{Multi-Access Edge Computing}
\acrodef{IoT}{Internet of Things}
\acrodef{DNN}{Deep Neural Network}
\acrodef{RL}{Reinforcement Learning}
\acrodef{PPO}{Proximal Policy Optimization}
\acrodef{DAG}{Directed Acyclic Graph}
\acrodef{SDN}{Software-Defined Networking}
\acrodef{NFV}{Network Function Visualization}
\acrodef{QoS}{Quality of Service}
\acrodef{MDP}{Markov Decision Process}
\acrodef{DL}{Deep Learning}
\acrodef{RAN}{Radio Access Network}
\acrodef{DRL}{Deep Reinforcement Learning}
\acrodef{TPTO}{Transformer-PPO based Task-Offloading}
\acrodef{BERT}{Bidirectional Encoder Representations from Transformers}
\acrodef{RNN}{Recurrent Neural Network}
\acrodef{LSTM}{Long Short Term Memory}
\acrodef{PaLM}{Pathways Language Model}
\acrodef{VM}{Virtual Machine}
\acrodef{GAE}{General Advantage Estimator}
\acrodef{UAV}{Unmanned Aerial Vehicles}
\acrodef{MIP}{Mixed Integer Programming}

\ifCLASSOPTIONcompsoc
  \usepackage[nocompress]{cite}
\else
  \usepackage{cite}
\fi

\begin{document}
\title{TPTO: A Transformer-PPO based Task Offloading Solution\\for Edge Computing Environments}

\author{
    \IEEEauthorblockN{
        Niloofar Gholipour\IEEEauthorrefmark{1},
        Marcos Dias de Assuncao\IEEEauthorrefmark{1},
        Pranav Agarwal\IEEEauthorrefmark{1},
        Julien Gascon-Samson\IEEEauthorrefmark{1},
        Rajkumar Buyya\IEEEauthorrefmark{2}
    }
    \IEEEauthorblockA{
        \IEEEauthorrefmark{1}Dept. of Software Engineering and IT, École de Technologie Supérieure, Univ. of Quebec, Montreal, Canada\\
        \{niloofar.gholipour, pranav.agarwal\}.1@ens.etsmtl.ca, \{marcos.dias-de-assuncao, julien.gascon-samson\}@etsmtl.ca
    }
    \IEEEauthorblockA{
        \IEEEauthorrefmark{2}CLOUDS Lab, School of Computing and Information Systems, The Univ. of Melbourne, Australia\\
        rbuyya@unimelb.edu.au
    }
}

\maketitle
\thispagestyle{plain}
\pagestyle{plain}

\begin{abstract}
Emerging applications in healthcare, autonomous vehicles, and wearable assistance require interactive and low-latency data analysis services. Unfortunately, cloud-centric architectures cannot fulfill the low-latency demands of these applications, as user devices are often distant from cloud data centers. Edge computing aims to reduce the latency by enabling processing tasks to be offloaded to resources located at the network's edge. However, determining which tasks must be offloaded to edge servers to reduce the latency of application requests is not trivial, especially if the tasks present dependencies. This paper proposes a \ac{DRL} approach called TPTO, which leverages Transformer Networks and \ac{PPO} to offload dependent tasks of IoT applications in edge computing. We consider users with various preferences, where devices can offload computation to an edge server via wireless channels. Performance evaluation results demonstrate that under fat application graphs, TPTO is more effective than state-of-the-art methods, such as Greedy, HEFT, and MRLCO, by reducing latency by 30.24\%, 29.61\%, and 12.41\%, respectively. In addition, TPTO presents a training time approximately 2.5 times faster than an existing DRL approach.

\end{abstract}

\IEEEpeerreviewmaketitle

\section{Introduction}
Edge computing, by complementing the cloud, can enable an increasing range of IoT applications that produce vast amounts of time-sensitive data requiring prompt analysis, such as in autonomous driving, healthcare, online video processing, and wearable assistance \cite{chen2021deep, yousefpour2019all}. In autonomous driving, for instance, latency is a critical factor in ensuring the safety of passengers and pedestrians. A minor delay in processing sensor data or making control decisions can not only degrade the users' quality of experience but also result in accidents or compromised safety. Edge computing provides computing services (e.g., base stations, access points, and edge routers) that are closer to end-users, contributing to lower the latency of application requests, their energy consumption, and the amount of data transferred to the cloud for processing \cite{kirkpatrick2013software}. 

Reducing the latency of IoT applications requires offloading data processing tasks to edge servers, an activity that often poses significant challenges. Offloading tasks can free constrained resources of user devices, but on the other hand, transferring data between the user devices and remote edge computing servers can impact the application latency \cite{tong2020adaptive}. Moreover, according to research conducted by Alibaba, around $75\%$ of real-world applications have interdependent tasks, commonly structured as a \ac{DAG}, where the vertices represent data sources, data sinks, end-users, and operators, and the edges represent data streaming from one operator to another \cite{goudarzi2020application,souza2020scalable}. Trying to devise efficient offloading decisions for these applications can often result in NP-hard problems, which require sophisticated algorithms to address them effectively.

Several heuristics, meta-heuristics, and model-based approaches exist for offloading in edge computing, most of which are unsuitable to stochastic environments where resource availability is continuously evolving \cite{wang2020fast,goudarzi2021distributed}. Edge computing is also stochastic when considering the number of applications, the number of tasks in an application, their arrival rate, their dependencies, and their resource requirements \cite{cao2019intelligent}. \ac{DRL} with policy optimization is a promising approach to address these challenges and design agents interacting with the environment to learn an optimal policy, enhanced over time through trial and error \cite{arulkumaran2017brief}. \ac{DRL} agents can learn a stochastic policy without having preliminary information about the environment, making them suitable for stochastic and complex systems like edge environments \cite{wang2020fast,goudarzi2021distributed,faraji2022self,hashem2022advanced,zheng2022deep}. 

We formulate the task offloading decision as a binary optimization problem and propose a solution, \ac{TPTO}, which utilizes a combination of \ac{MDP}, \ac{RL}, and Transformers \cite{vaswani2017attention}. While \ac{RL} provides a learning mechanism to optimize offloading decisions over time, the Transformer model enhances the solution's performance by enabling it to learn from previous tasks and apply the knowledge to future offloading decisions. \ac{TPTO} trains Transformers for various edge computing tasks and quickly adapts to new ones with less training time and shorter latency. Our proposed approach features \ac{BERT} architecture incorporating multi-head attention, layer normalization, and feed-forward fully connected layers. The predictions made by the Transformer, provided to a Softmax function, act as the actions that guide the training process in collaboration with the \ac{PPO} algorithm. This results in a more efficient and effective solution. To our knowledge, this paper presents the first work that applies \ac{BERT} for offloading decisions in edge environments. To validate our approach, we carry out simulations using synthetic \acp{DAG} that reflect real-world applications with dependent tasks and network topologies with multiple wireless transmission rates. Experimental results demonstrate our approach's effectiveness in optimizing the offloading problem.

The main contributions of this work are: A novel latency-aware task offloading approach, \ac{TPTO}, that leverages the Transformer model that quickly adapts to stochastic edge environments; and a new policy that jointly uses Transformers and an actor-critic framework to determine the best action for task offloading -- i.e., offloaded to the edge or processed locally to minimize end-to-end latency. 

The paper is structured as follows: Section \ref{sec:problem} describes the problem and presents a formulation. Section \ref{sec:solution} presents \ac{TPTO}, whereas Section \ref{sec:evaluation} analyzes and compares its efficiency against state-of-the-art techniques. Section \ref{sec:related_work} reviews related work, and Section \ref{sec:conclusion} concludes the paper and discusses future work.


\section{Problem Description and Formulation}
\label{sec:problem}

\begin{figure}[htb]
\centering
\includegraphics[width=1.\columnwidth]{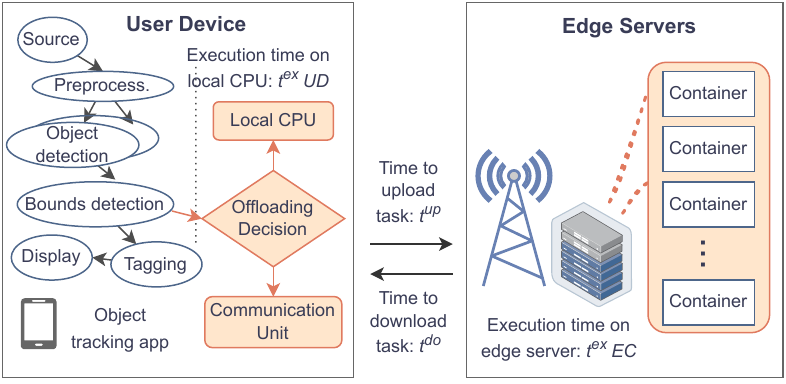}
\caption{System model with sample application.}
\label{fig:system_model}
 \end{figure}

A real-time object detection system presents a typical example of an application that can benefit from computation offloading to edge computing servers (Figure~\ref{fig:system_model}). In this scenario, a user device often captures a video stream from a camera and aims to detect and recognize objects from the video feed in real-time. This scenario reflects, for instance, applications in facial recognition \cite{wang2020fast} and pest bird detection systems \cite{Mahmud2022CoPI}. The user device can carry out data pre-processing and execute a lightweight object detection model locally (e.g., identifying some features), but the type of computations it can perform will largely depend on the system status, the available resources, and their constraints. Alternatively, some of the computations can be offloaded to an edge server. 

An application $A$ is a \ac{DAG} $G = (V, E)$ where each vertex $v_i \in V$ represents a task and each directed edge $e(v_i, v_j) \in E$ is a dependence constraint in which task $v_i$ must complete before task $v_j$ starts. \textit{Entry tasks} are tasks without parent tasks, whereas \textit{exit tasks} or \textit{sinks} are tasks without children. The computation of task $v_i$ corresponds to the number of CPU cycles needed for its execution, given by $c_i$. Moreover, we define as $\textrm{data}^{up}_i$ and $\textrm{data}^{do}_i$ the amount of data required to upload and download, respectively, task $v_i$ to/from an edge server. 

The computing capacity of a resource $m_j$ (user device or edge server), denoted as $cs_j$, reflects its clock speed times the number of cores available in the system. A user device is associated with a container providing the computing and network resources that an application requires. We consider that the containers share the computing resources equally, such that the capacity of a container on an edge server $m_j$ is $cs_{ct} = cs_j / k$, where $k$ is the number of users in $m_j$. This approach of equal resource allocation ensures a fair distribution of the computing capacity to each user, thereby promoting efficient utilization of the resources within the edge computing environment.
The user device can execute a task locally or offload its computation to an edge server via \textit{wireless channels}. A wireless channel's uplink and downlink transmission rates are $r_{up}$ and $r_{do}$. Three steps are required to offload a task $v_i$ to an edge server $m_j$: first, the user device sends the task to the edge server via a wireless channel. Second, the edge server executes the task. Finally, the edge server sends the execution results back to the user's device. 
The overall latency for a task is influenced by both the task's requirements and the current system status. Hence, the total time involved in offloading task $v_i$ to edge server $m_j$ encompasses the time to upload the task ($t^{up}_{i}$), the time to execute the task on the edge server ($t^{ex}_{i}$), and the time to download the resulting data back to the user device ($t^{do}_{i}$). This can be mathematically expressed as:
\begin{equation}
\label{eq:latencies}
\small t^{up}_{i} = data^{up}_i / r_{up}, \quad t^{ex}_{i} = c_i / cs_{ct}, \quad t^{do}_{i} = data^{do}_i / r_{do}
\end{equation}

When offloaded to an edge server, the overall end-to-end latency of task $v_i$ represents the sum of the above times in (\ref{eq:latencies}). On the other hand, if a user device executes task $v_i$ locally, hence using resource $m_k$ (the user device), its latency consists only of the task execution time (\textit{i.e.} $t^{ex}_{i} = c_i / cs_{k}$). In addition, for a task $v_i$ scheduled for execution, we establish four task finish times, namely $FT_{i}^{ud}$, $FT_{i}^{up}$, $FT_{i}^{ec}$, and $FT_{i}^{do}$, to denote the task finish time on the user device, on the upload link, on the edge server and the download link. If task $v_i$ runs locally on the user device, then $FT_{i}^{up} = FT_{i}^{ec} = FT_{i}^{do} = 0$. Otherwise, $FT_{i}^{ud} = 0$ if $v_i$ is offloaded to an edge server. 

Before scheduling a task $v_i$, all preceding tasks (\textit{i.e.}, its parent tasks) must have been scheduled. In this way, we denote $RT_{i}^{ud}$, $RT_{i}^{up}$, $RT_{i}^{ec}$, and $RT_{i}^{do}$ as the ready time, the earliest time that task $v_i$ can be executed on a resource (user device, upload link, edge server, download link) so that the precedence constraints are maintained. Hence, for task $v_i$ scheduled on the user device, we can calculate its ready time as:
\begin{equation}
\label{eq:rt_mobile}
\small  RT_{i}^{ud} = \max_{j\in parent(v_i)} max\left\{FT_{j}^{ud},FT_{j}^{do}\right\}
\end{equation}

\noindent where \textbf{\textit{parent}}$(v_i)$ is the set of parent tasks immediately before task $v_i$. $RT_{i}^{ud}$ is the earliest time at which all the tasks preceding $v_i$ will have completed and produced the results that $v_i$ requires. When a task $v_j$ preceding $v_i$ is scheduled locally, then $max\{FT_{j}^{ud},FT_{j}^{do}\} = FT_{j}^{ud}$; otherwise, when offloaded to the edge server, $max\{FT_{j}^{ud},FT_{j}^{do}\} = FT_{j}^{do}$. Task $v_i$ can only start executing once $v_j$ has freed the wireless download channel. 

On the other hand, if that task $v_i$ is to be offloaded to the edge server, then its ready time on the upload channel ($RT_{i}^{up}$) is given by:
\begin{equation}
\label{eq:rt_up}
\small RT_{i}^{up} = \max_{j\in parent(v_i)} max\left\{FT_{j}^{ud},FT_{j}^{up}\right\}
\end{equation}

\noindent where $RT_{i}^{up}$ is the earliest time when $v_i$ can use the upload channel while meeting precedence constraints. When a task $v_j$ preceding $v_i$ is scheduled locally, then $max\{FT_{j}^{ud},FT_{j}^{up}\} = FT_{j}^{ud}$; otherwise, when offloaded to the edge server, then $max\{FT_{j}^{ud},FT_{j}^{up}\} = FT_{j}^{up}$. Task $v_i$ can only start execution once $v_j$ has freed the wireless download channel.

The ready time of a task $v_i$ on an edge server is:
\begin{equation}
\label{eq:rt_mec}
\small RT_{i}^{ec} = max\left\{FT_{i}^{up}, \max_{j\in parent(v_i)} FT_{j}^{ec}\right\}
\end{equation}

\noindent where $RT_{i}^{ec}$ is the earliest time $v_i$ can execute on the edge server while respecting precedence constraints. If a task $v_j$ preceding $v_i$ is scheduled locally, then $FT_{j}^{ec} = 0$. Hence, $\max_{j\in parent(v_i)} FT_{j}^{ec}$ is the earliest time when all offloaded tasks preceding $v_i$ have finished execution. 

The earliest time for sending the results of task $v_i$ back to the user device is: 
\begin{equation}
\label{eq:rt_do}
\small RT_{i}^{do} = FT_{i}^{ec}
\end{equation}

The offloading goal is to compute an offloading plan $O_n = (o_1, o_2, \ldots, o_n)$ that minimizes the latency of an application DAG $G(V, E)$, where $n = |V|$. Here, $o_i$ denotes the offloading decision for task $v_i$, where $o_i$ can be either $0$ for local computation or $1$ for remote computation.
Before offloading, tasks are sorted by priority, as discussed later, so that $O_{n-1}$, for example, represents the partial offloading plan comprising all tasks from $v_1$ to $v_{n-1}$.   
The optimization goal is, hence, to minimize the overall \textit{Application Latency}:
\begin{equation}
\label{eq:total_latency}
\small AL_{O_n} = max \left[ max_{v_e \in \mathcal{E}}(FT_e^{ud}, FT_e^{do}) \right]
\end{equation}

\noindent where $\mathcal{E}$ is the set of exit tasks (\textit{i.e.} tasks with no children). The equation considers the maximum task latency within a DAG to compute the overall application latency. This maximum time represents the duration of the critical path of the DAG, which is the longest path from a start task to any of the exit tasks. Table \ref{tab:HParam} summarizes the main notations used in this paper.

\newcommand{\wdesc}{.68\columnwidth}
\newcommand{\wnot}{.2\columnwidth}
\begin{table}[htb]
\centering
\scriptsize
\label{Notation}
\caption{Notation used in this paper.}
\begin{tabular}{cp{\wdesc}}
\toprule
\textbf{Notation} & \textbf{Description} \\ 
\toprule
\multirow{2}{*}{$G(V,E)$}  &  Application DAG where $V$ is the set of tasks and $E$  \\
& the task precedence constraints \\
\cmidrule(rl){1-2}
$v_i \in V$  &  Computing task $v_i$ \\
\cmidrule(rl){1-2}
\multirow{2}{*}{$e(v_j, v_i) \in E$}  &  Precedence constraint, task $v_j$ must execute before  \\
& $v_i$ can start \\
\cmidrule(rl){1-2}
\multirow{2}{*}{$data_{i}^{up}$, $data_{i}^{do}$}   &  Number of bytes to upload/download to/from an edge \\ 
&  server when offloading task $v_i$ \\ 
\cmidrule(rl){1-2}
$r_{up}$, $r_{do}$  &   Transmission rates of wireless uplink and downlink channels     \\ 
\cmidrule(rl){1-2}
$cs_k$, $cs_{ct}$   &    Computing capacity of resource $m_k$, and of a container    \\ 
\cmidrule(rl){1-2}
\multirow{2}{*}{\textbf{$t_{i}^{up}$, $t_{i}^{ex}$, $t_{i}^{do}$}}   &   Time required for uploading, executing and \\ 
& downloading task $v_i$ to edge server $m_k$ \\
\cmidrule(rl){1-2}
\multirow{2}{\wnot}{\centering\textbf{$FT_{i}^{ud}$, $FT_{i}^{up}$, $FT_{i}^{ec}$, $FT_{i}^{do}$}} & Finish time of task ${v_i}$ on user device, uplink channel, \\ 
& edge server, and downlink channel \\
\cmidrule(rl){1-2}
\multirow{2}{\wnot}{\centering\textbf{$RT_{i}^{ud}$, $RT_{i}^{up}$, $RT_{i}^{ec}$, $RT_{i}^{do}$}} & Earliest time when task $v_i$ can use the user device, \\
& uplink channel, edge server, and downlink channel \\
\bottomrule
\end{tabular}
\end{table}

\section{Transformer-Based Offloading Solution}
\label{sec:solution}

This section presents our Transformer-PPO-based task offloading solution.

\subsection{Transformer-PPO based Task Offloading}
\label{subsec:TPTOmodel}

In \ac{RL}, an agent interacts with an environment, trying to learn a policy to take actions that maximize the accumulated reward. An \ac{MDP}, commonly used to represent \ac{RL} problems \cite{Sutton:2018RL}, consists of a tuple $(S, A, P, R, \gamma)$, where $S$ represents the set of possible states; $A$ represents the action space; $P(s'|s, a)$ denotes the probability of transitioning to state $s'$ when taking action $a$ under the current state $s$; $R(s, a, s')$ represents the immediate reward received when transitioning from $s$ to $s'$ by taking action $a$; $\gamma$ is a discount factor. The goal is to find a policy $\pi(s)$ that maximizes the expected cumulative reward over time. A policy network $\pi(a|s, \theta)$ takes the state $s$ as input and outputs a probability distribution over the actions $a$, where $\theta$ represents the neural network parameters. Training the policy network involves finding the optimal parameters $\theta^{*}$ that maximize the expected cumulative reward, a process typically performed using policy gradient algorithms that seek to maximize the expected return. \ac{TPTO} optimizes the policy network parameters using \ac{PPO} \cite{schulman2017proximal}. During training, \ac{PPO} uses a batch of sampled trajectories to update the network weights. The following describes the main elements of our \ac{MDP}:

\textbf{State $S$}: A state comprises the task profile (CPU cycle requirements and data sizes), the DAG topologies, the wireless transmission rates, and the status of edge resources. The status of an edge resource depends on the offloading decisions for tasks preceding $v_i$. Hence, we can express the state combining the encoded DAG and the partial offloading plan as:
\begin{equation}
\small S = \left\{s_i | s_i =(G (V, E), O_{i})\right\}
\end{equation}

\noindent where $i \in [1, |V|]$, $G(V, E)$ represents the sequence of embedding tasks and $O_{i}$ is the partial offloading plan of task $v_i$. We use the approach outlined in \cite{wang2020fast} to convert a DAG into a sequence of embedding tasks. 

For efficient offloading, tasks receive a "rank" based on their completion time and dependencies, sequenced from lowest to highest rank. Each task is embedded with information on its attributes and its parent-child relationships. This approach ensures optimized task scheduling, enhancing efficiency and reducing latency. Task embeddings use three vectors: one for the task's profile, one for parent tasks, and another for child tasks, with padding if the number of tasks is below the vector's length.

\textbf{Action $A$}: As the scheduling for each task is a binary choice, executing the task either on the user device or on an edge server, the action space is $A := {0, 1}$, where $0$ represents execution on the user device and, $1$ represents offloading.

\textbf{Reward function $R$}: The objective is to minimize the total application latency, defined in Equation \ref{eq:total_latency}. Hence, the reward function estimates the negative increase in latency resulting from an offloading decision for a particular task: $\Delta AL_{O_i} = AL_{O_i} - AL_{O_{i-1}} $, where $AL_{O_i}$ represents the total latency when taking a given action for task $v_i$ and $AL_{O_{i-1}}$ represents the total latency of the partial offloading plan for the previous task.

Assume that $ \pi (a_{i} | G(V, E), O_{i-1})$ represents the likelihood of the offloading plan $O_{i-1}$ given the graph $G(V, E)$, we can compute $\pi(O_n | G(V, E)) $ by using the chain rule of probability on each $ \pi (a_{i} | O_{i-1}, G(V, E))$ as follows:
\begin{equation}
 \label{eq:policy}
    \small  \pi(O_{n} | G(V, E)) = \prod_{i=1}^n \pi (a_{i}|O_{i-1}, G(V,E))
 \end{equation}

We employ Transformers to devise our policy. Distinct from traditional \acp{RNN}, Transformers use an encoder-decoder structure, effectively addressing various RNN limitations. Typically, the encoder integrates features like embedding, multi-head attention, residual connections with normalization, feed-forward networks, and softmax. A distinguishing feature of Transformers is the incorporation of a self-attention mechanism, enhancing data dependency extraction \cite{vaswani2017attention}. In the context of \ac{TPTO}, a Transformer processes the task embeddings from a sequence $(v_1, v_2, ..., v_n)$ of a DAG and formulates a refined representation through successive Transformer layers. Based on the output, the actor makes offloading decisions for each task. Meanwhile, the critic assesses each task's value function, with fully connected layers producing these outputs.

\subsection{Implementing TPTO}

\begin{figure}[htb]
    \begin {center}
    \includegraphics[width=1.\columnwidth]{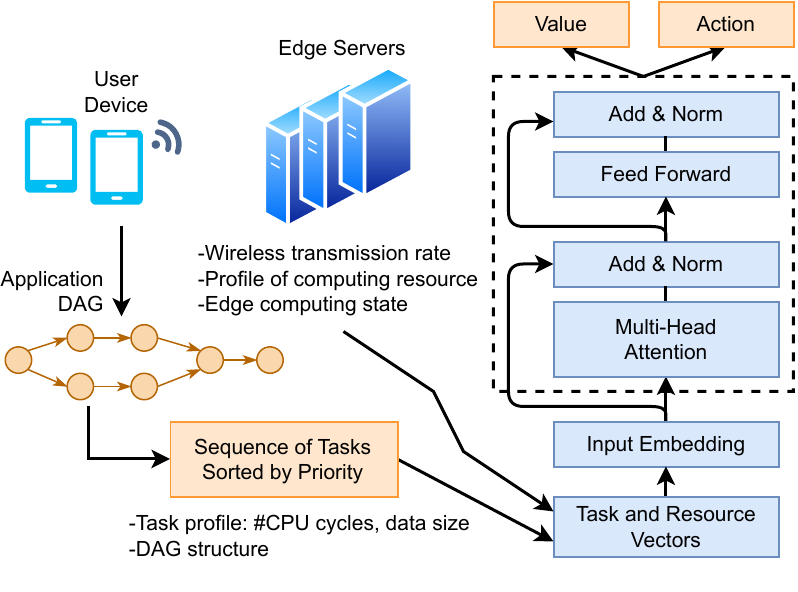}
    \caption{Overview of TPTO.}
    \label{fig:tpto_overview}
    \end {center}
    \end{figure}

As Figure \ref{fig:tpto_overview} outlines, TPTO employs the Transformer model and \ac{PPO} to update the policy network. First, the Transformer receives an observation of the environment and produces two results: the policy logits and the value function. The policy logits are passed through a softmax function to obtain a proper probability distribution of the available actions. Next, the actor network takes the Transformer's output and produces the final policy, which provides a probability distribution for the available actions. Finally, the critic network takes the Transformer's output and generates the estimated value of the current state. The advantage function captures the difference between the actual and estimated return and the estimated value of the current state.

We use \ac{PPO} as the policy optimization method. For a given learning task $\mathcal{T}$, \ac{PPO} creates trajectories using a sample policy $\pi_{\theta_{sam}}$ and updates the target policy $\pi_{\theta}$ over multiple epochs, where $\theta$ and $\theta_{sam}$ are the parameters of the target and sample policies, respectively. At the initial epoch, ${\theta} = \theta_{sam}$. Then the probability ratio $r_t(\theta)$ at a time step $t$ is:
\begin{equation}
\label{eq:prob_ratio}
\small r_t(\theta) = \frac{\pi_{\theta}(a_t | s_t)}{\pi_{\theta_{sam}}(a_t | s_t)}
\end{equation}

\noindent where $s_t = G(V, E), O_t$. To update the actor's policy, \ac{PPO} uses a clipped surrogate objective to avoid extensive policy updates:
\begin{equation}
\label{eq:ppo_clip_obj}
\small L^{CLIP}(\theta) = \hat{\mathbb{E}}_t\left[min(r_t(\theta)\hat{A}_t, clip(r_t(\theta),1-\epsilon, 1+ \epsilon)\hat{A}_t) \right]
\end{equation}

\noindent where $\hat{A}_t$ is the advantage function at time step $t$, and $\hat{\mathbb{E}}$ is the average expectation over a set of samples in an algorithm that alternates between sampling and optimization \cite{schulman2017proximal}. As the policy and value functions share most of their parameters, facilitating mutual training, we also employ the entropy coefficient to compute the entropy bonus, added to the policy loss, to encourage exploration in the policy space. The combined objective is, therefore:
\begin{equation}
\label{eq:ppo_comb_obj}
\small L^{CLIP+VF+S}(\theta) = \hat{\mathbb{E}}_t\left[L_t^{CLIP}(\theta) - c_{1}L_t^{VF}(\theta) + c_{2}S[\pi_\theta](s_t) \right]
\end{equation}

\noindent where $c_1$ and $c_2$ are coefficients, $S[\pi_\theta](s_t)$ represents the entropy bonus, and $L_t^{VF}(\theta)$ is the squared-error loss: $(V_\theta(s_t) - V_t^{targ})^2$, where $V$ is a state-value function.

The advantage function at time step $t$, denoted by $\hat{A}_t$, is calculated using \ac{GAE} \cite{schulman2015high}. GAE is a specific type of advantage function estimated as follows:
\begin{equation}
\label{eq:ppo_advantage}
\small \hat{A}_t =\sum_{l=0}^{n-t+1}(\gamma\lambda)^k \left[r_t + \gamma V(s_{t+k+1}) - V(s_{t+k})\right]
\end{equation}

\noindent where $\lambda$ is in the interval $(0, 1)$ and determines the equation's balance between bias and variance. We can then use gradient ascent to maximize $L^{CLIP+VF+S}(\theta)$.

\begin{algorithm}[htb]
\caption{Transformer-PPO based task offloading}
\label{alg:TPTO}
\begin{algorithmic}[1]
\begin{small}
\Require Task distribution $r(\mathcal{T})$, learning rate $\alpha$
\Ensure Updated policy parameters $\theta$

\State Randomly initialize the parameters of the policy, $\theta$;

\For{iterations $k \in \{1, 2, \ldots, K\}$}
    \State Sample $n$ learning tasks $\{ \mathcal{T}_0, \mathcal{T}_1, \ldots, \mathcal{T}_n \}$ from $r(\mathcal{T})$;
    
    \For{each task $\mathcal{T}_i$}
        \State Initialize $\theta_{\text{sam}} \leftarrow\theta$
        \State Sample trajectory set $S = (\tau_0, \tau_1, \ldots, \tau_n)$ from $\mathcal{T}_i$ using policy $\pi(\theta_{\text{sam}})$;
        \State Calculate the advantage estimates $\hat{A}_1, \hat{A}_2,..,\hat{A}_T$;
\State Compute the policy gradient:
        \State $L_{\tau_{\text{sam}}}^{\text{TPTO}}(\theta_{\text{sam}}) = \nabla_{\theta_{\text{sam}}} L^{\text{CLIP+VF+S}}(\theta_{\text{sam}})$
    \EndFor
   \State Update the policy network parameters $\theta$ using Adagrad optimizer with gradients computed by the TPTO loss function with trajectory set $S$ for $m$ steps:
    \State $\theta \gets \theta + \alpha L_{\tau_{\text{sam}}}^{\text{TPTO}}(\theta_{\text{sam}})$
\EndFor
\end{small}
\end{algorithmic}
\end{algorithm}

Algorithm~\ref{alg:TPTO} outlines how TPTO performs the offloading decision and generates trajectories. First, the algorithm samples an $n$ sized batch of learning tasks $\tau$ and performs the training loop for each sampled learning task. Following the completion of the training loop, the algorithm then updates the policy parameters $\theta$ using gradient ascent $\theta\leftarrow\theta + \alpha L^{TPTO}$ using Adagrad optimizer \cite{kingma2014adam}, where $\alpha$ is the learning rate of training loop.

\section{Performance Evaluation}
\label{sec:evaluation}

This section presents the experimental setup, the baseline algorithms, and performance evaluation results.

\subsection{Experimental Setup}

We evaluated the performance of TPTO by developing an event-driven simulation environment in Python using the OpenAIGym \cite{brockman2016openai}, similar to \cite{wang2020fast}. This approach ensured a controllable and repeatable evaluation process. We consider a cellular network whose data transmission rate varies based on the user devices' position. Also, a user device's CPU clock speed is $1 GHz$, denoted by $f_1$. In contrast, each container in an edge server has a quota of four cores, each core running at $2.5 GHz$, represented by $f_s$. Consequently, offloaded tasks can simultaneously use all cores, resulting in a combined CPU clock speed of $10 GHz$ for each container. 

\begin{figure}[htb]
\centering
\includegraphics[width=.95\columnwidth]{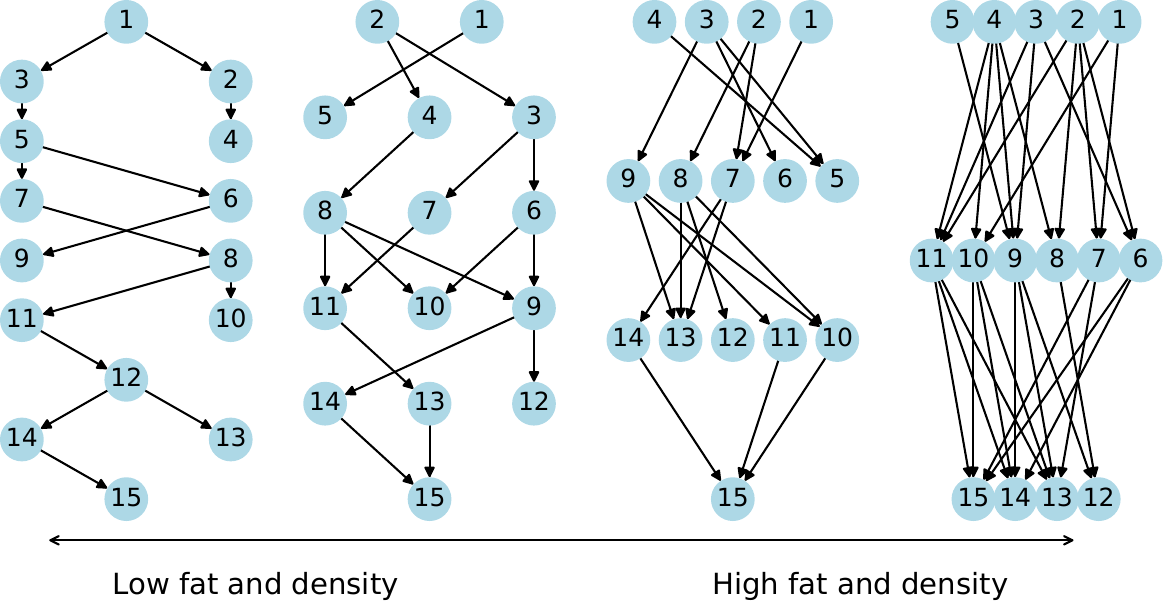}
\caption{Examples of application DAGs with different fats and densities.}
\label{fig:dagmodel}
 \end{figure}

 We consider latency under multiple scenarios to evaluate TPTO's efficiency comprehensively in dynamic environments. We use a synthetic DAG generator tool\footnote{\url{https://github.com/frs69wq/daggen}} to create diverse heterogeneous DAGs that emulate real-world applications. These DAGs encompass a broad spectrum of topologies and transmission rates encountered in practical scenarios.
 The generator receives four parameters: $n$, $fat$, $density$, and $ccr$. The $n$  represents the number of tasks; $fat$ determines the DAG's width and height; $density$ sets the number of edges between two levels of the DAG; and computation to communication ratio, $ccr$, specifies the ratio between tasks' communication and computation cost. 

\begin{table}[hbt]
\centering
\scriptsize
\caption{TPTO's hyperparameters.}
\label{tab:HParam}
\begin{tabular}{p{40mm}c}
\toprule
\centering\textbf{Hyperparameter}       & \textbf{Value}  \\ 
\toprule
Number of Layers     &   3  \\ 
Num Attention Head   &   8  \\ 
Dimension of Key Vector   &  1024 \\ 
Dimension of Value Vector &  1024 \\ 
Dimension of FF network &  512  \\ 
Hidden Size          &   512 \\ 
Dropout Rate         &   0.4 \\ 
Policy Learning Rate     &  0.1 \\ 
Valuefunc Learning Rate  &  0.01 \\ 
Batch Size           &   100  \\ 
Clip ratio           &   0.2  \\ 
Activation Function  &   Relu \\ 
Optimization Method  &   Adagrad \\ 
Discount Factor      &   0.99  \\ 
Entropy coefficient  &   0.5   \\ 
\bottomrule
\end{tabular}
\end{table}

To model the mobile network users' diverse preferences, we generated $25$ DAG datasets, each dataset comprising $100$ DAGs with various fat, densities, and $ccr$ -- key parameters impacting the DAG topology. Each DAG has $20$ tasks, and fat and density values for each DAG are selected randomly from the set $\{0.4, 0.5, 0.6, 0.7, 0.8\}$, while $ccr$ is chosen randomly within the range of $0.3$ to $0.5$. This range is representative of the computation sensitivity observed in a majority of IoT applications. The DAGs simulate diverse application preferences of a mobile user: for instance, a fatter DAG suggests a preference for more parallel tasks, while a denser DAG indicates a higher dependency between tasks, all under varying data transmission speeds.
We randomly select $22$ DAG sets as ``training datasets'' and the remaining three as ``unseen testing datasets'' with different DAG topologies. Figure \ref{fig:dagmodel} illustrates DAGs generated by the synthetic DAG generator with varying fat and density. 

TPTO is implemented using Tensorflow, with $3$ layers of Transformer encoders having $512$ hidden units per layer and layer normalization included. Table~\ref{tab:HParam} summarizes the hyperparameters for training TPTO. To ensure the robustness of the TPTO policy, we trained it using a range of transmission rates between 4Mbps to 22Mbps, with a step size of 3Mbps. To evaluate its performance on previously unseen transmission rates and topologies, we tested the trained policy on data rates of 8.5Mbps and 11.5Mbps, not seen during training, following a similar methodology as in \cite{wang2020fast} with sampling 20 trajectories for a DAG on the dataset. In addition, as we aim to assess how TPTO performs in different dynamic scenarios, the task data size varies from 5KB to 50KB, while the CPU cycle requirements range from $10^{7}$ to $10^{8}$ cycles per task, as reported in \cite{dinh2017offloading}. Furthermore, the length of the parent/child task indices vector is 12. By testing TPTO's performance on these diverse sets of DAGs, we aim to gain insights into its ability to effectively provision network resources and meet the varying needs of mobile users.

\subsection{Baseline Algorithms}

We assess TPTO's performance against three state-of-the-art algorithms:

\textbf{MRLCO}: 
this algorithm, proposed by Wang \textit{et al.} \cite{wang2020fast},  integrates meta reinforcement learning and a Seq2Seq neural network. The approach focuses on modeling task offloading using meta-reinforcement learning and an offloading policy based on a custom Seq2Seq neural network.

\textbf{HEFT based:}  this algorithm, based on the work by Lin \textit{et al.} \cite{lin2014task}, involves prioritizing tasks using the HEFT method and scheduling each task according to its earliest estimated finish time. 

\textbf{Greedy Heuristic:} a greedy approach considers the estimated finish time of each task to decide whether to assign a task to the user device or an edge server.

\subsection{Result Analysis}

\begin{figure*}[htbp]
	\centering
    \subfigure[~TPTO \ \label{fig:tpto}]
 {\includegraphics[width=.37\textwidth]{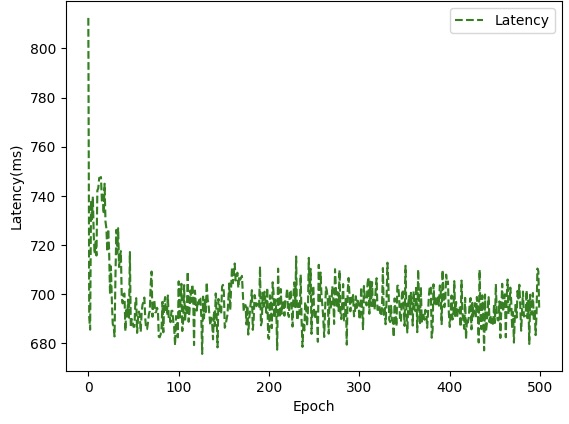}}\,
  \subfigure[~MRLCO\label{fig:mrlco}]
 {\includegraphics[width=.37\textwidth]{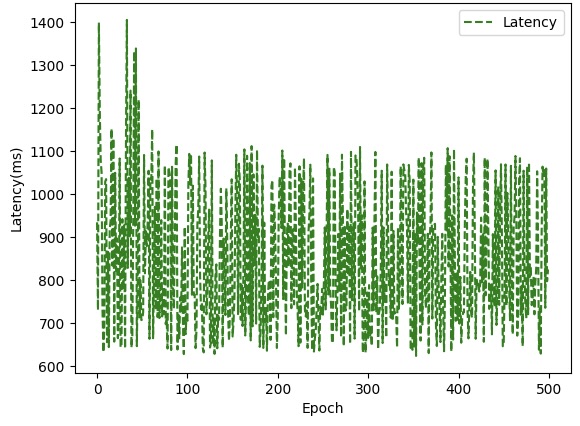}}\,
	\subfigure[~HEFT\label{fig:heft}]{\includegraphics[width=.37\textwidth]{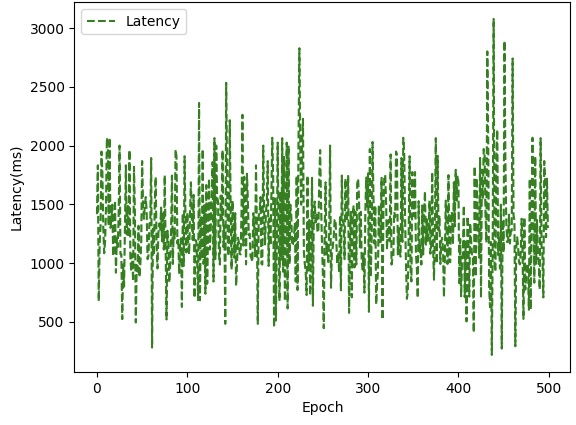}}
	\subfigure[~Greedy\label{fig:greedy}]{\includegraphics[width=.37\textwidth]{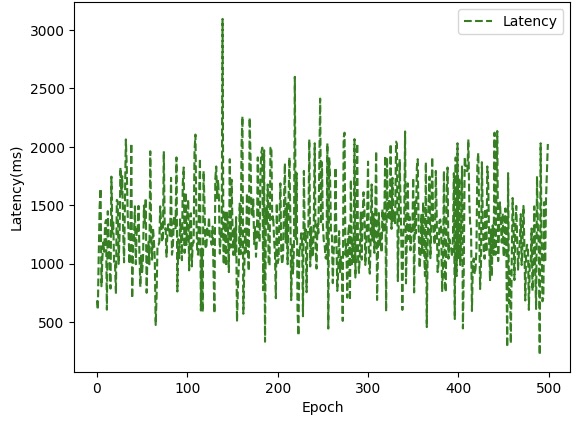}}\,
	\caption{Influence of wireless transmission rate and network topology on latency.} 
	\label{fig:latency_all}
\end{figure*}

Figures \ref{fig:tpto} and \ref{fig:mrlco} depict the average latency of simulation results during training for TPTO and MRLCO. The results demonstrate that TPTO converges faster than MRLCO while being more stable and general, mainly due to TPTO's ability to effectively capture the diverse preferences of mobile users through its training on a wide range of network topologies and transmission rates. Figure \ref{fig:heft} and \ref{fig:greedy} show the performance of HEFT and Greedy algorithms.

\begin{table}
\caption{Comparative analysis of TPTO and baseline methods: average latency (ms) across diverse test datasets.}
\label{tab:comparison}
\centering
\footnotesize
\begin{tabular}{c@{\hskip 3mm}c@{\hskip 3mm}c@{\hskip 5mm}c}
\toprule
\multirow{4}{15mm}{\centering\textbf{Testing Topology Sets}} & \multirow{4}{*}{\textbf{Algorithm}} & \multicolumn{2}{c}{\textbf{Wireless Transmission Rate}}\\ 
& & \multicolumn{2}{c}{\textbf{of $r_{up}$ and $r_{do}$}} \\
\cmidrule(){3-4} 
& & \textbf{8.5Mbps} & \textbf{11.5Mbps} \\ 
\midrule
\multirow{4}{*}{\begin{tabular}[c]{@{}c@{}} \textbf{fat = 0.8}\\ \textbf{density = 0.6}\\ \textbf{ccr = 0.5}\end{tabular}} & \textbf{HEFT}   & 1033 & 835  \\  
& \textbf{Greedy}  & 1064 & 837  \\ 
& \textbf{MRLCO} & 846 & 760   \\  
& \textbf{TPTO}   & 741 & 581\\ 
\midrule
\multirow{4}{*}{\begin{tabular}[c]{@{}c@{}} \textbf{fat = 0.5}\\ \textbf{density = 0.7}\\ \textbf{ccr = 0.3}\end{tabular}} &\textbf{HEFT}   & 1157 & 849   \\  
& \textbf{Greedy}   & 1462   & 952 \\ 
& \textbf{MRLCO}  & 989 & 869   \\  
& \textbf{TPTO}   & 1022 & 811  \\ 
\midrule
\multirow{4}{*}{\begin{tabular}[c]{@{}c@{}} \textbf{fat = 0.6}\\ \textbf{density = 0.8}\\ \textbf{ccr = 0.4}\end{tabular}}  & \textbf{HEFT}   & 1521 & 943   \\  
& \textbf{Greedy}  & 1009 & 822   \\ 
& \textbf{MRLCO}  & 894 & 810   \\  
& \textbf{TPTO}  & 900 & 719  \\ 
\bottomrule
\end{tabular}
\end{table}

Table \ref{tab:comparison} summarizes the average latency of TPTO and the baseline algorithms. TPTO outperforms heuristic and meta-learning algorithms for the various wireless transmission rates. Overall, the Greedy algorithm has the highest latency, while TPTO achieves lower latency under various network conditions, indicating its effectiveness in provisioning network resources to meet the needs of mobile users. Moreover, distinct topologies reflect the diverse preferences of user requests in terms of dependency and parallel computing of tasks. Increasing the transmission rate can further reduce latency as offloaded tasks traverse the wireless channels faster. Overall, the results show that TPTO is a promising solution for optimizing network performance and enhancing user experience in mobile networks. TPTO achieves a training time 2.5 faster than MRLCO. The Transformer architecture of TPTO is mainly responsible for this training time difference. Transformers are known for their parallel execution and efficient utilization of self-attention mechanisms, which can exploit the parallel processing capabilities of modern hardware architectures, resulting in a faster training process. These results underscore the potential benefits of employing Transformer-based models for optimizing offloading decisions in the edge computing environment.

\section{Related Work}
\label{sec:related_work}

Approaches for computation offloading to edge computing are a very active topic. Proposed techniques fall mainly into machine-learning and optimization-based methods.

\textbf{Machine-Learning Offloading Approaches:}
Qu \textit{et al.} present a framework for IoT devices to offload computing tasks to edge servers~\cite{qu2021dmro}. They use deep meta-reinforcement learning to minimize energy consumption, task computation, and transmission delays by dividing applications into sequential workflows. The proposed framework, Deep Meta Reinforcement learning based Offloading (DMRO), includes an inner and outer loop. The former relies on Q-learning, whereas the latter employs a meta-algorithm to learn the initial parameters and adapt to changing environments, quickly converging to optimal offloading solutions. 

Huang \textit{et al.}~\cite{huang2020meta} propose MELO, a Meta-Learning-based computation Offloading algorithm for independent tasks in edge computing, which consists of one edge server and $N$ wireless devices, each with a prioritized task to execute. They apply binary offloading, where tasks run locally on a device or the edge server. The approach focuses on minimizing latency, communication, and computation delay. 

\newcommand{\etal}{\textit{et al.\xspace}}
\begin{table*}[hbt]
\centering
\caption{Summary of related work and their comparison with TPTO.}
\label{relatedwork}
\begin{tabular}{cccccc}
\toprule
\multicolumn{1}{c}{\multirow{3}{*}{\textbf{Techniques}}} & 
\multirow{3}{*}{\textbf{Category}}                 & 
\multicolumn{1}{c}{\multirow{3}{16mm}{\centering\textbf{Task\\Dependency}}} & 
\multicolumn{3}{c}{\textbf{Task Offloading Engine}} \\ 
\cmidrule(rl){4-6} 
\multicolumn{3}{c}{} & \multirow{2}{*}{\textbf{Main Approach}} & \multirow{2}{10mm}{\centering\textbf{Task\\ Priority}} & \multirow{2}{18mm}{\centering\textbf{Network\\Architecture}} \\
\multicolumn{3}{c}{} & & \\
\toprule
Nguyen \etal~\cite{nguyen2023dependency} & \multirow{3}{*}{Optimization} & \cmark  & \multicolumn{1}{c}{Discrete whale optimization} & \multicolumn{1}{c}{\xmark} & \xmark  \\  
Abbas \etal~\cite{abbas2021meta} &   &  \xmark  & \multicolumn{1}{c}{Ant colony, whale, grey wolf}   & \multicolumn{1}{c}{\xmark}  & \xmark  \\  
Xu \etal~\cite{xu2020meta} &  & \xmark  & \multicolumn{1}{c}{Order-preserving policy, bisection search} & \multicolumn{1}{c}{\xmark }  &  \xmark \\ 
\midrule
Qu \etal~\cite{qu2021dmro} & \multirow{4}{*}{Machine learning} & \cmark & \multicolumn{1}{c}{Deep Meta Learning, Q-learning} & \multicolumn{1}{c}{\xmark} &  DNN \\  
Huang \etal~\cite{huang2020meta} & &  \xmark & \multicolumn{1}{c}{Meta-Learning} & \multicolumn{1}{c}{\cmark} &  DNN \\  
Yang \etal~\cite{yang2022deep} & &  \xmark & \multicolumn{1}{c}{Deep Supervised Learning} & \multicolumn{1}{c}{\cmark} &  CNN, DNN \\  
\multicolumn{1}{c}{\textbf{TPTO}}  &  & \multicolumn{1}{c}{\textbf{\cmark }}   & \multicolumn{1}{c}{PPO, Actor-Critic} & \multicolumn{1}{c}{\cmark} & Transformer NN \\ 
\bottomrule
\end{tabular}
\end{table*}

Yang \textit{et al.} \cite{yang2022deep} tackle joint offloading optimization and bandwidth allocation, modeled as a mixed-integer programming (MIP) problem for independent tasks. They propose the Deep Supervised Learning-based computational Offloading (DSLO) algorithm that considers task delay and energy consumption. Furthermore, incorporating batch normalization into two classical neural network architectures, CNN and DNN, enhances the convergence speed of DSLO.

\textbf{Optimization-based Offloading Techniques:}
Nguyen \textit{et al.} \cite{nguyen2023dependency} introduce a collaborative scheme for \acp{UAV} to share workloads. They consider the task topology, which involves splitting a task into sub-tasks with dependencies and the power consumption constraints of the \acp{UAV} in edge computing. They use the discrete whale optimization algorithm and CVXPY's SCS solver to solve the optimization problem, modeled as a mixed-integer, non-linear, and non-convex problem. 
Abbas \textit{et al.} \cite{abbas2021meta} present classical approaches for optimal independent task offloading in edge computing environments. They use well-known meta-heuristics such as the ant colony optimization algorithm, whale optimization algorithm, and Grey wolf optimization algorithm, adapting these algorithms to their problem. The goal is to minimize the energy consumption of user devices and IoT and minimize response time for task computation in edge computing. 
A search-based meta-heuristic model, introduced by Xu \textit{et al.} \cite{xu2020meta}, also focused on task offloading and time allocation in edge computing for independent tasks. Considering computation rate and task execution latency, they formulated the problem as a \ac{MIP} and divided it into sub-problems: offloading decision and resource allocation. They proposed an ``order-preserving policy generation method'', which works well in large networks.

Machine learning approaches often demonstrate superior performance than traditional optimization methods. Still, existing work generally employs conventional DRL with sequential neural networks, resulting in less computation efficiency and extended training time. Moreover, most real-world applications comprise dependent tasks, which prior work generally ignores \cite{xu2020meta}, \cite{abbas2021meta}, \cite{yang2022deep}, \cite{huang2020meta}. TPTO tackles these limitations, exhibiting fast adaptability to new tasks by applying Transformers and effectively minimizing latency -- an essential concern for delay-sensitive applications -- by strategically considering task dependencies. Table \ref{relatedwork} summarizes and compares related works with TPTO.

\section{Conclusions and Future Work}
\label{sec:conclusion}

This work introduced TPTO, a distributed DRL method for task offloading optimization in edge computing using Transformers to reduce latency in DAG-structured applications. We first introduced a latency model that optimizes the task execution time, communication, and offloading in an edge computing environment. This model serves as the basis for the decision-making process in TPTO. Then, experimental results demonstrated TPTO's effectiveness under various network conditions and topologies. TPTO presents superior performance compared to three baseline algorithms: MRLCO, HEFT, and Greedy. In addition, TPTO consistently achieved the lowest latency, showcasing its ability to make efficient offloading decisions. Future research will focus on TPTO's scalability in larger edge computing setups and explore multi-criteria optimization, including energy consumption and execution cost.

\balance
\bibliographystyle{IEEEtran}
\bibliography{references}
\end{document}